\documentclass[%
 reprint,
superscriptaddress,
 amsmath,amssymb,
 aps,
prb,
]{revtex4-2}

\usepackage{calrsfs}
\usepackage{graphicx}% Include figure files
\usepackage{dcolumn}% Align table columns on decimal point
\usepackage{bm}% bold math
\usepackage{braket}
\usepackage[dvipsnames]{xcolor}
\usepackage{xcolor}
\usepackage{amsmath}% http://ctan.org/pkg/amsmath
\usepackage{chemformula} % Formula subscripts using \ch{}
\usepackage[T1]{fontenc} % Use modern font encodings
\usepackage{graphics}
\usepackage{makecell}
\usepackage{xcolor}
\usepackage{colortbl}
\usepackage{amsmath}
\usepackage{enumerate}
\usepackage[normalem]{ulem}
\usepackage{multirow}
\usepackage[bottom]{footmisc}
\usepackage{bm}
\usepackage{siunitx}
\usepackage{multirow}
\usepackage{listings}

\usepackage{forest}

\definecolor{folderbg}{RGB}{124,166,198}
\definecolor{folderborder}{RGB}{110,144,169}

\def\Size{4pt}
\tikzset{
  folder/.pic={
    \filldraw[draw=folderborder,top color=folderbg!50,bottom color=folderbg]
      (-1.05*\Size,0.2\Size+5pt) rectangle ++(.75*\Size,-0.2\Size-5pt);  
    \filldraw[draw=folderborder,top color=folderbg!50,bottom color=folderbg]
      (-1.15*\Size,-\Size) rectangle (1.15*\Size,\Size);
  }
}

\begin{document}
%%%%%%%%%%%%%%%%%%%%%%%%%%%%%%%%%%%%%%%%%%%%%%%%%%%%%%%%%%%%%%%%%%%%%
\newcommand*\mycommand[1]{\texttt{\emph{#1}}}
\newcommand*\rd[1]{{\color{red} #1}}
\newcommand*\bl[1]{{\color{blue} #1}}
\newcommand\mathplus{+}
\newcommand{\snrxn}[0]{S$_\mathrm{N}$2}
\newcommand{\erxn}[0]{E2}
\newcommand{\Ea}[0]{$E_\mathrm{a}$}
\newcommand{\dEa}[0]{$\Delta E_\mathrm{a}$}
\newcommand{\Eas}[0]{$E_\mathrm{a}^{\mathrm{S}}$}
\newcommand{\Eae}[0]{$E_\mathrm{a}^{\mathrm{E}}$}
\newcommand{\deltaml}[0]{$\Delta$\nobreakdash-ML}

\definecolor{codegreen}{rgb}{0,0.6,0}
\definecolor{codegray}{rgb}{0.5,0.5,0.5}
\definecolor{codepurple}{rgb}{0.58,0,0.82}
\definecolor{backcolour}{rgb}{0.95,0.95,0.92}

\lstdefinestyle{mystyle}{
    backgroundcolor=\color{backcolour},   
    commentstyle=\color{codegreen},
    keywordstyle=\color{magenta},
    numberstyle=\tiny\color{codegray},
    stringstyle=\color{codepurple},
    basicstyle=\ttfamily\footnotesize,
    breakatwhitespace=false,         
    breaklines=true,                 
    captionpos=b,                    
    keepspaces=true,                 
    %numbers=left,                    
    numbersep=5pt,                  
    showspaces=false,                
    showstringspaces=false,
    showtabs=false,                  
    tabsize=2
}

\lstset{style=mystyle}

%\title{Geometry Relaxation and Transition State Search with Quantum Machine Learning} 
%\title{Fast and accurate equilibrium and transition state geometries using operator quantum machine learning in chemical compound space}
\title{Reducing Training Data Needs with Minimal Multilevel Machine Learning (M3L)} 

\author{Stefan Heinen}
\affiliation{Vector Institute for Artificial Intelligence, Toronto, ON, M5S 1M1, Canada}
\author{Danish Khan}
\affiliation{Vector Institute for Artificial Intelligence, Toronto, ON, M5S 1M1, Canada}
\affiliation{Department of Chemistry, University of Toronto, St. George Campus, Toronto, ON, Canada}
\author{Guido Falk von Rudorff}
\affiliation{University Kassel, Department of Chemistry, Heinrich-Plett-Str.40, 34132 Kassel, Germany}
\affiliation{Center for Interdisciplinary Nanostructure Science and Technology (CINSaT), Heinrich-Plett-Straße 40, 34132 Kassel}
\author{Konstantin Karandashev}
\affiliation{University of Vienna, Faculty of Physics, Kolingasse 14-16, AT-1090 Wien, Austria}
\author{Daniel Jose Arismendi Arrieta}
\affiliation{Department of Chemistry-{\AA}ngstr\"om Laboratory, Uppsala University, Box 538, SE-75121 Uppsala, Sweden}
\author{Alastair J. A. Price}
\affiliation{Acceleration Consortium, University of Toronto. 80 St George St, Toronto, ON M5S 3H6}
\affiliation{Departments of Chemistry, University of Toronto, St. George Campus, Toronto, ON, Canada}
\author{Surajit Nandi}
\affiliation{Department of Energy Conversion and Storage, DTU, Anker Engelunds Vej, DK-2800 Kgs. Lyngby}
\author{Arghya Bhowmik}
\affiliation{Department of Energy Conversion and Storage, DTU, Anker Engelunds Vej, DK-2800 Kgs. Lyngby}
\author{Kersti Hermansson}
\affiliation{Department of Chemistry-{\AA}ngstr\"om Laboratory, Uppsala University, Box 538, SE-75121 Uppsala, Sweden}
\author{O. Anatole von Lilienfeld}
\email{anatole.vonlilienfeld@utoronto.ca}
\affiliation{Vector Institute for Artificial Intelligence, Toronto, ON, M5S 1M1, Canada}
\affiliation{Acceleration Consortium, University of Toronto. 80 St George St, Toronto, ON M5S 3H6}
\affiliation{Department of Materials Science and Engineering, University of Toronto, St. George campus, Toronto, ON, Canada}
\affiliation{Department of Chemistry, University of Toronto, St. George campus, Toronto, ON, Canada}
\affiliation{Department of Physics, University of Toronto, St. George campus, Toronto, ON, Canada}
\affiliation{Machine Learning Group, Technische Universität Berlin and Berlin Institute for the Foundations of Learning and Data, Berlin, Germany}

\begin{abstract}
For many machine learning applications in science, data acquisition, not training, is the bottleneck even when avoiding experiments and relying on computation and simulation. 
Correspondingly, and in order to reduce cost and carbon footprint,  training data efficiency is key.  
We introduce minimal multilevel machine learning (M3L) which 
optimizes training data set sizes using a loss function at multiple levels of reference data in order to minimize a combination of prediction error with overall training data acquisition costs (as measured by computational wall-times).
Numerical evidence has been obtained for calculated atomization energies and electron affinities of thousands of organic molecules at various levels of theory including HF, MP2, DLPNO-CCSD(T), DFHFCABS, PNOMP2F12, and PNOCCSD(T)F12, and treating tens  with 
basis sets TZ, cc-pVTZ, and AVTZ-F12.
Our M3L benchmarks for reaching chemical accuracy in distinct chemical compound sub-spaces 
indicate substantial computational cost reductions by factors of 
$\sim$ 1.01, 1.1, 3.8, 13.8 and 25.8 when compared to heuristic sub-optimal multilevel machine learning (M2L) for the data sets QM7b, QM9$^\mathrm{LCCSD(T)}$, EGP, QM9$^\mathrm{CCSD(T)}_\mathrm{AE}$, and QM9$^\mathrm{CCSD(T)}_\mathrm{EA}$, respectively.
Furthermore, we use M2L to investigate the performance for 76 density functionals when used within multilevel learning and building on the following levels drawn from the hierarchy of Jacobs Ladder:~LDA, GGA, mGGA, and hybrid functionals.
Within M2L and the molecules considered, 
mGGAs do not provide any noticeable advantage over GGAs. 
Among the functionals considered and in combination with LDA, the three on average top performing GGA and Hybrid levels for atomization energies on QM9 using M3L correspond respectively to PW91, KT2, B97D, and $\tau$-HCTH, B3LYP$\ast$(VWN5), TPSSH. 
\end{abstract}
\maketitle
\section{Introduction}
%TODO Konstantin: need to check all abbreviations being written in full when introduced (GPU, ANN, etc.)
%
% Big Challenge: what remains unsolved
Machine learning (ML) has revolutionized various scientific fields by leveraging large data sets to extract valuable insights.
In 2012 Rupp \textit{et al.} \cite{Rupp2012} introduced the first ML approach successfully learning atomization energies.
However, the acquisition cost of training data remains a substantial bottleneck, often hindering progress in tackling new challenges.
Even more so, data generation in general is still in its infancy: Most of all data has only been collected in the past two decades\cite{AlJarrah2015}.

Additionally, we emphasize the significance of computational atomistic simulations, consuming approximately 35\% of High-Performance Computing (HPC) budgets, which highlights the urgent need for more efficient methodologies\cite{Heinen2020}.
Recognizing the growing importance of sustainability, we also mention the environmental implications of using data-inefficient ML models; emphasizing the drive to make science greener. 
This latter point has become increasingly crucial in light of growing electricity prices.
Hence, our research centers on minimizing training data acquisition cost and economizing data generation.
Through the utilization of multilevel machine learning (M2L), we have refined the training set sizes using established legacy optimization techniques such as Bayesian optimization resulting in minimal multilevel machine learning (M3L).
This strategic approach yields notable reduction in training acquisition time, all the while upholding model accuracy.
Therefore, this work has the potential to greatly expedite the process of materials discovery by relieving the financial constraints linked to computational resources.
The cost limitation, a frequent bottleneck for MAP optimizers, will be effectively eliminated through our proposed workflow.
\\\\
%when computational chemistry and materials use $\sim$0.5$\cdot 10^3$ Peta Flops on the top 500 Super Computers\cite{Heinen2020}.
% maybe remove (not necessary/important)
%Already in 1996, a least square fit of parameters in graph based component-wise run time estimates in parallelized self consistent field computations of atoms  was introduced by Papay et al \cite{PAPAY1996}.
%Such predictive models may even comprise direct minimization of the estimated environmental impact of a calculation as the target quantity in the model\cite{Garg2011}.
%Furthermore, a potentially valuable application in the context of quantum chemistry may be the run time optimization of a given tensor contraction scheme on a specific hardware by predictive modeling techniques\cite{8445573}.
Numerous researchers have already explored the realm of mitigating computational costs: Heinen \textit{et al.}\cite{Heinen2020} in learning run times of quantum chemical (QC) calculations and using those predictions to optimize scheduling on super computers.
Duan \textit{et al.}\cite{Duan2019} machine learned the success rate of transition metal geometry optimization calculations and therefore were able to stop potentially not converging calculations and free up space.
Showing the importance of performance and cost in quantum chemistry and artificial intelligence (AI) for interatomic potentials is shown in Review\cite{Zuo2020}.
This noteworthy review shows the impact of the computational discovery of materials in the era of big data\cite{Lu2021}.
% add more delta/multi learning (P. Zaspel 2023)
Most recently, considerable amount of work has been invested in better algorithms such as central processing unit (GPU) enhanced density functional theory (DFT) codes\cite{epifanovsky2021software, Andrade2021}.
Furthermore, new improved representations like many body distribution functionals (MBDF)\cite{mbdf} among others\cite{amons_slatm,fchl18,fchl19,soap2018} and new neural network infrastructures\cite{schnet,schnetpack,painn,allegro} minimizing the computational cost of either scaling linearly with system size or reducing the need for larger training data to achieve chemical accuracy (1kcal/mol).
For the benchmark set QM9\cite{qm9}, modern representations in Kernel Ridge Regression (KRR) and Artificial Neural Networks (ANN) structure achieve chemical accuracy at 3000, 4000, 8000, 9000 training data for FCHL18\cite{fchl18}, Allegro\cite{allegro}, FCHL19/PaiNN\cite{fchl19,painn}, and SLATM\cite{amons_slatm}, respectively.
MBDF reaches Chemical accuracy at 10k, however, scaling linearly with size - therefore being compact - drastically reduces the training and prediction time\cite{mbdf}.
\begin{figure*}[!ht]
\includegraphics[width=0.99\textwidth]{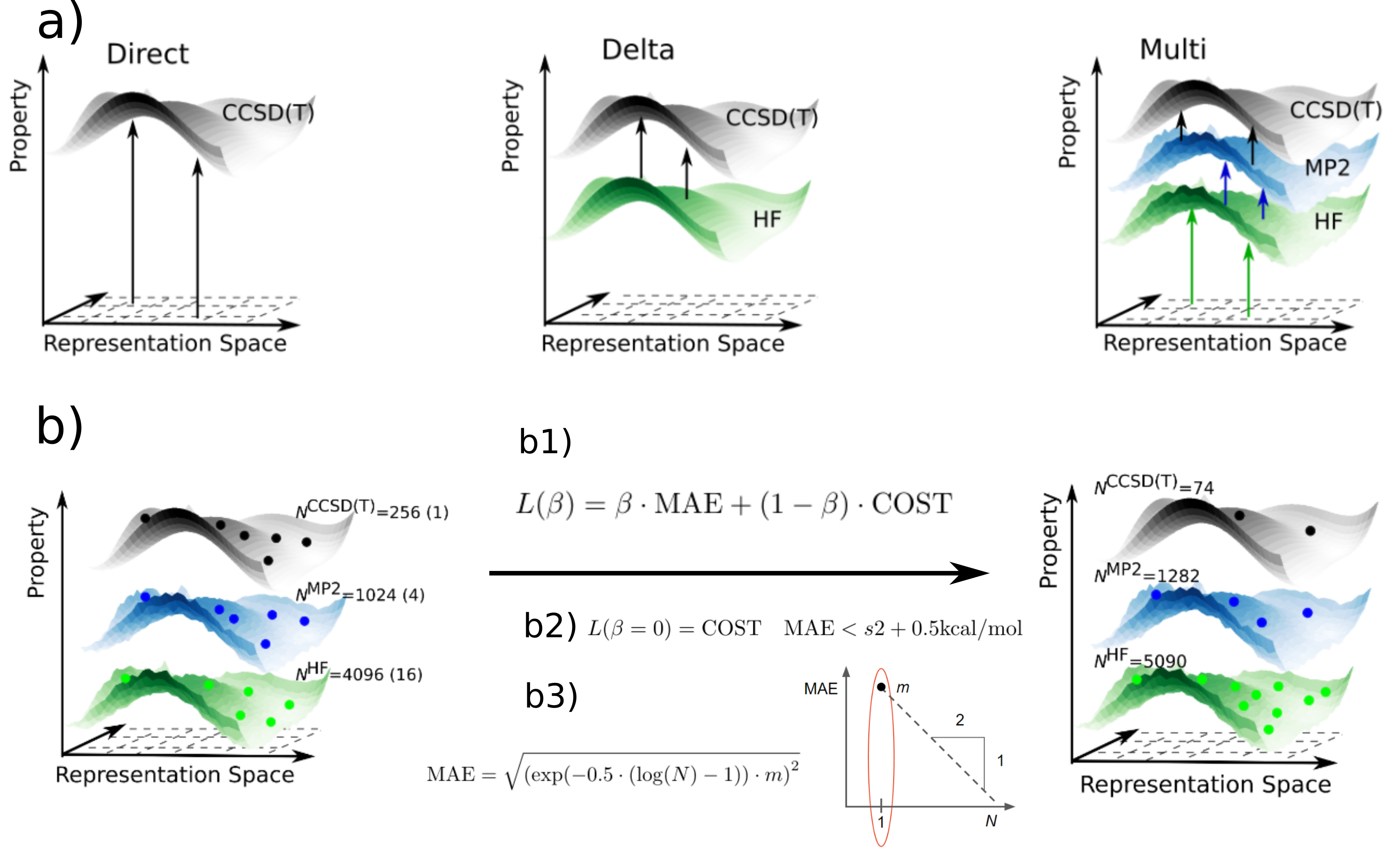}
\label{fig:intro}
\caption{(a) Fictitious potential energy surfaces for HF, MP2, and CCSD(T) showing the concept of direct, $\Delta$, and multilevel learning, respectively. Arrows denote ML predictions, smooth surfaces calculated values (Back Surface target property (CCSD(T))) and coarse surfaces denote ML predictions. (b) Fictitious potential energy surfaces showing the training set sizes of the original work (s2=1;4;16) left and after Bayesian optimization (b1 and b2) with cost functions $L(\beta)$ and $L$ with constraint $s2 < \mathrm{MAE}$ or an educated guess using learning theory (b3) right, respectively. Using loss function in b2 for QM7b dataset yields new ratios right hand size on b) showing a reduction from 256 to 74 CCSD(T) training data.}
\end{figure*}
% Some transfer learning and computational cost state of the art SWITCH DELTA AND TRANSFER LEARNING (use Fig 1 to explain)
A different approach is to pre-train a neural network on a lower level of theory and subsequently (using transfer learning) training the ANN on the higher level of theory, reducing the training cost\cite{Jha2019,Yamada2019, Lentelink2020}.
Another noteworthy approach is to use a cheaper base line ($\Delta$ learning, Figure \ref{fig:intro} a)) and predict the difference between the base line and the target level of theory.
Already in 2015 multilevel learning was applied to protein-protein interactions using ANN\cite{Zubek2015}.
Simultaneously, Ramakrishnan \textit{et al.}\cite{Ramakrishnan2015} introduced $\Delta$ML using a composite strategy that adds machine learning corrections to computationally inexpensive approximate legacy quantum methods to efficiently screen chemical compound space\cite{vonLilienfeld2020}.
%TODO <arghya>: <can we mention, simple linear corrections have been explored also https://pubs.acs.org/doi/full/10.1021/acs.jcim.2c00760 >)
By adding a base line they were able to shift the learning curve to the left meaning a lower off set through adding a lower level of theory and therefore reducing the need for high level of theory training data.
Most recently, Goodlett \textit{et al.} compared different multi fidelity approaches for potential energy surface predictions\cite{Goodlett2023}, showing $\Delta$ learning as the most favorable approach.
A drawback of this method is the need for a base line calculation on the lower level for each  prediction.
The main differences between $\Delta$- and multilevel learning is that (a) the lowest level (base line) is learned not calculated, and (b) that one can benefit from having multiple levels of theory in the data (as is shown in Figure \ref{fig:intro}).
Building upon previous work\cite{Zaspel2018}, we introduce Minimal Multilevel Machine Learning (M3L), we highlight the limitations of the previous heuristic-based multilevel machine learning (M2L) techniques, which lacked optimization.
%The heuristics approaches used were ratios of $2^0, 2^1, 2^2$ (1:2:4) and $2^0, 2^2, 2^4$ (1:4:16) named $s1$ and $s2$, respectively.
%The better preforming ratios ($s2$) were used in this work as a null model method to compare to.

% This work
M3L overcome this limitation by enhancing cost efficiency through systematic optimization of training set sizes at each level. 
To establish the efficacy of M3L, we have performed extensive numerical experimentation using datasets consisting of small organic molecules on various levels of theory, including Hartree-Fock (HF), MP2, Coupled Cluster Singles Doubles perturbative triples (CCSD(T)) on various basis sets (TZ, cc-pVTZ, and AVTZ-F12). 
We leverage Bayesian optimization to reduce the training set sizes required for higher levels of theory as shown in Figure \ref{fig:intro} b1) and b2).
Furthermore, we employ learning theory and learning curves\cite{vapnik1994learningcurves} to make educated estimates of the training set sizes (Figure \ref{fig:intro} b3).
By applying these approaches, we successfully reduce the amount of high-level theory data needed while maintaining the desired accuracy.
We provide numerical evidence showcasing the potential cost reductions offered by quantum mechanical data, highlighting their advantages compared to direct (conventional) quantum machine learning (QML), M2L and M3L approaches saving up to 25 times training data acquisition cost.

The algorithms proposed in this work can also be extended to evaluate the usefulness of intermediate steps of density functionals, the working horse in quantum machine learning\cite{Huang2023}, in Jacobs Ladder.
We demonstrate that by considering 76 DFT functionals on Jacob's ladder and showing hat there are no discernible benefits of using mGGAs (meta-generalized gradient approximations) over GGAs (generalized gradient approximations) within this hierarchy.

\begin{figure*}[ht!]
\includegraphics[width=0.89\textwidth]{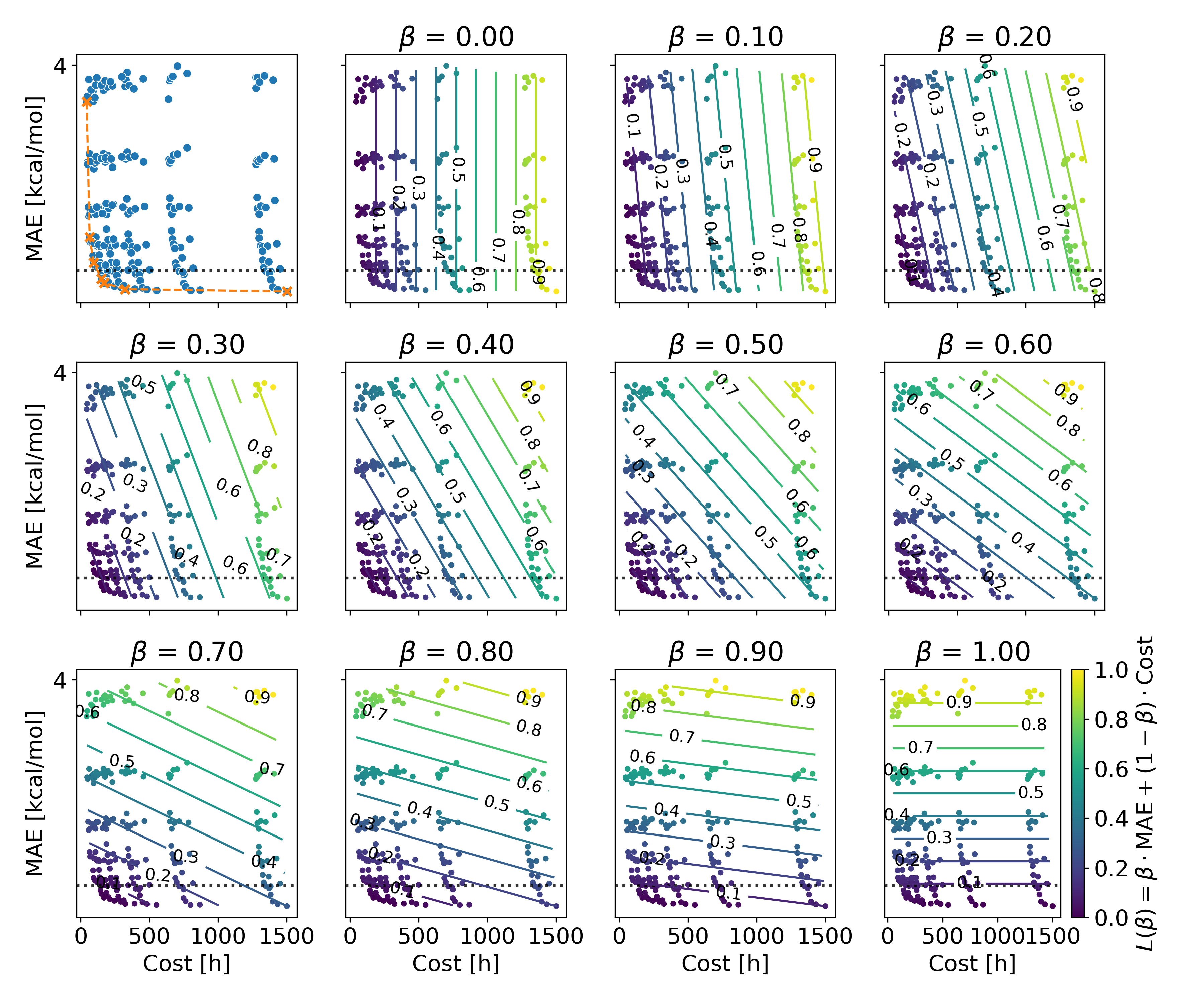}
\caption{Results of grid scan ($N_i$ = [175, 375, 750, 1500, 3000, 6000] for $i=\mathrm{HF, MP2, CCSD(T)}$) shown for different values of $\beta$ of cost function $L(\beta) = \beta \cdot \mathrm{MAE}/\mathrm{MAE}^\mathrm{max} + (1-\beta)\cdot \mathrm{Cost}/\mathrm{Cost}^\mathrm{max}$ for the QM7b data set. Top left: Pareto front (orange) for the lowest values for $L(\beta)$ for each $\beta$}
\end{figure*}

\section{Methods}
\subsection{Kernel Ridge Regression}

KRR\cite{Vapnik1998,gpr_rasmussen} was employed for all machine learning tasks using the QMLcode framework\cite{qmlcode2017}.
A detailed derivation can be found in Ref.\cite{Heinen2020}.
To train a model, the regression coefficients $\mathbf{\alpha}$ are calculated using the closed-form solution given by
\begin{equation}
\mathbf{\alpha} = \left( \mathbf{K}^{\mathrm{train}} + \lambda\mathbf{I}\right)^{-1}\mathbf{y}^{\mathrm{train}}
\label{eq:train}
\end{equation}
where $\mathbf{K}^{\mathrm{train}}$ represents the training kernel, $\lambda$ is the regularization strength which was not used in this work and $\mathbf{I}$ is the identity matrix.
Lambda is only necessary when there is numerical noise associated with the data or kernel inversion procedures for larger training set sizes or if there is considerable amount of noise present in the data.
Since neither were a factor in this work we set it to zero.
For each Kernel element, a gaussian kernel was used %KONSTANTIN Laplacian?
\begin{equation}
k(\mathbf{x}_i , \mathbf{x}_q) = \exp\left( -\frac{||\mathbf{x}_i - \mathbf{x}_q||^2_2}{2\sigma^2} \right),
\label{eq:gaussian}
\end{equation}
where $k(\mathbf{x}_i , \mathbf{x}_q)$ is a Laplacian kernel element\cite{Deisenroth2020,anatolebook}.  %KONSTANTIN no need to continue the sentense?
To make ML based estimates $y_q$ for a query compound $\mathbf{x}_q$, the following equation is used
\begin{equation}
y_q(\mathbf{x}_q) = \sum_i^N \alpha_i\ k(\mathbf{x}_i , \mathbf{x}_q).
\label{eq:KRR}
\end{equation}
Here, $\mathbf{x}$ represents the representations, and $\sigma$ is a hyperparameter. The hyperparameter $\sigma$ was optimized using $k$-fold cross-validation, as described in\cite{Heinen2020}.
The representation used throughout this work was MBDF from\cite{mbdf}.
Since it is a compact presentation (scales linearly with system size), it allows for Bayesian optimizations by calling predictions multiple times for different training set sizes.

\subsection{Multi-Level Learning}
Zaspel \textit{et al.} were the first to demonstrate the applicability of  multilevel machine learning in terms of a grid-combination technique in 2018.\cite{Zaspel2018}. 
Ratios between the different levels were chosen heuristically as $s1 = 2^0, 2^1, 2^2$ and $s2 = 2^0, 2^2, 2^4$ for the training set sizes $N^{\mathrm{tot}} = N^{\mathrm{CCSD(T)}} : N^{\mathrm{MP2}} : N^{\mathrm{HF}}$.
The better preforming ratios $s2$ from M2L were used in this work as a null model method to compare to.
The ML based estimate of property $y$ is as follows:

\begin{equation}
y^{\mathrm{CCSD(T)}}_q(\mathbf{x}_q) = y_q^{\mathrm{HF}}(\mathbf{x}_q) + y_q^{\mathrm{HF}\rightarrow\mathrm{MP2}}(\mathbf{x}_q) + y_q^{\mathrm{MP2}\rightarrow\mathrm{CCSD(T)}}(\mathbf{x}_q)
\end{equation}
As shown in Figure \ref{fig:intro} a)  direct learning is performed for the lowest level ($y^{\mathrm{HF}}_q(\mathbf{x}_q)$), while differences are learnt between the low-mid (HF $\rightarrow$ MP2) and mid-high (MP2 $\rightarrow$ CCSD(T)) levels. 
As evinced in that figure and as obvious from 
Eq.~(\ref{eq:KRR}), the sum does not necessarily run over the same training molecules, implying that there is no need to obtain labels at all levels for any given training molecule. 

\subsection{Data Sets}
The following three datasets were used in this study:
\begin{itemize}
    \item \textbf{QM7b}\cite{qm7b, blum}: This dataset includes 7,211 molecules with a maximum of 7 heavy atoms.
    The molecules contain elements such as C, N, O, S or Cl. 
    Based on the string representation (SMILES\cite{Weininger1988,Weininger1990}) of molecules in the database, the universal force field\cite{rappe} was used to generate reasonable 3D geometries,
    as implemented in OpenBabel\cite{OBoyle2011}.
    The resulting geometries were relaxed using the PBE\cite{Perdew1996} level of theory and tier2 basis sets in the FHI-aims\cite{FHI-aims} electronic structure code. 
    \item \textbf{QM9}\cite{qm9,Ruddigkeit2012}: This dataset comprises 130k small organic molecules including C, N, O, F atoms. Geometries were optimized at the B3LYP/6-31G(2df,p)\cite{B3LYP_functional,LYP} level of theory.
    \item \textbf{EGP}: The Electrolyte Genome Project (EGP)\cite{Qu_Persson:2015} contains ionization potential (IP) and electron affinities (EA) of 4830 molecules relevant to battery electrolytes at the B3LYP/6-31+G$\ast$ level of theory.
\end{itemize}

which were divided into the five following subsets:
\begin{itemize}
\item \textbf{QM7b (toy set)}: On the QM7b DFT geometries Zaspel \textit{et al.}\cite{Zaspel2018} performed HF\cite{Hartree1928,Slater1928}, MP2\cite{Frisch1990,HeadGordon1988}, and CCSD(T) Single Point calculations using three different basis sets (sto-3g\cite{Hehre1969,Collins1976}, 6-31G\cite{Ditchfield1971}, cc-pVTZ\cite{Kendall1992}). cc-pVTZ was used in this work. The single point calculations were performed using Gaussian09\cite{g09}.

\item \textbf{QM9$^\mathrm{LCCSD(T)}$}: For a subset of QM9 $\sim$18'000 HF, MP2, and local PNO-LCCSD(T) calculations were performed on a cc-pVTZ basis set using ORCA5\cite{Neese2022}.

\item \textbf{QM9$^\mathrm{CCSD(T)}_{\mathrm{AE/EA}}$}: This subset of the QM9 dataset includes atomization energies and electron affinities for $\sim$4'000 molecules on PNO-CCSD(T)-F12/AVTZ-F12, and PNO-MP2/AVTZ-F12, DF-HF-CABS levels of theory, respectively.
Electron affinities are composed of two calculation and the total run time of both was taken for this work.
AE and EA stand for atomization energies and electron affinities, respectively.
Another property in this data set are the ionization potentials.
However, HF predictions are significantly worse compared to AE and EA, which makes learning inefficient.
We excluded this data set from the main manuscript and put the learning curves in the SI.

\item \textbf{EGP}:  a subset was generated providing atomization energies containing 600 PNO-CCSDT-F12, 1'200 PNO-MP2, and 1'200 DF-HF-CABS levels of theory.
The subset contains up to 15 heavy atoms (B, C, N, O, F, Si, P, S, Cl, and Br) 92 in total.
%TODO <arghya>: <the dataset is called MultiXC-QM9 and the DOIs are https://doi.org/10.26434/chemrxiv-2022-fs70n-v2 and https://doi.org/10.11583/DTU.c.6185986>). 
\item \textbf{QM9$^\mathrm{DFT}$}\cite{Nandi2022}: This dataset includes all 133k molecules in the QM9 dataset.
For each molecule, atomization energies were calculated using 76 different functionals and three basis sets (SZ, DZ, TZ).
The triple zeta basis set (TZ) was specifically used in this study.
\end{itemize}

\subsection{Computational Cost \& Loss Function}
To combine both Mean Absolute Error (MAE) and CPU times (COST), a loss function ($L(\beta)$) was used. In this work, both properties were scaled between 0 and 1, and a weighted sum approach was employed: %KONSTANTIN: expand what "CPU" is.
\begin{equation}
L(\beta) = \beta \cdot \mathrm{MAE} + (1 - \beta) \cdot \mathrm{COST}
\label{eq:cost}
\end{equation}
where $0 \le \beta \le 1$ represents the weight for MAE and the cost, respectively. The MAE is defined as:

\begin{equation}
\mathrm{MAE} = \frac{\sum_q^{N} |y_{q,\mathrm{est}} - y_{q,\mathrm{ref}}|}{N} / \mathrm{error}^{\mathrm{max}}
\end{equation}

and the COST is defined as:

\begin{equation}
\mathrm{COST} = \frac{\left[ \sum_i^{N^\mathrm{CCSD(T)}} t^{\mathrm{CCSD(T)}}_i + \sum_j^{N^\mathrm{MP2}} t^{\mathrm{MP2}}_j + \sum_k^{N^\mathrm{HF}} t^{\mathrm{HF}}_k \right]}{\mathrm{COST}^\mathrm{max}}
\end{equation}
As the Gaussian electronic structure code\cite{g09} does not provide timing information, the approach described in\cite{Heinen2020} was used to predict CPU times for the calculations, estimating the cost of training set generation for the QM7b toy set.
To predict the timings, the standard representation and KRR models were employed, substituting the property $\mathbf{y}^{\mathrm{train}}$ with CPU times instead of atomization energies.
For all other data sets, we obtained the real/calculated timings from the output files, more details can be found in the SI.
Figure 2 shows a grid scan for the QM7b data set for a range of training set sizes of $N_i = [175, 375, 750, 1500, 3000, 6000]$ for $i$ being HF, MP2, CCSD(T), respectively.
For every data point the loss function was evaluated for different values of $\beta$ and color encoded in Figure 2.
On the top left subplot the MAE were plotted in blue and the lowest value for each loss function was marked in orange, showing a Pareto front, indicating the validity of the weighted sum approach.
\subsection{Bayesian Optimization}
The Bayesian optimization as implemented in the scikit-optimize python library (skopt)\cite{scikit-learn} was used for optimizing the training set sizes in all datasets.
This global integer optimization allowed for a broad scan of the training set sizes using:
i) the loss function $L(\beta)$ (Eq. \ref{eq:cost} and Figure \ref{fig:intro} b1)) for different hyperparameters $\beta = [0.25, 0.5, 0.75]$,
ii) using the same loss function but setting $\beta$ = 0  giving $L(\beta=0) = \mathrm{COST}$ with the constraint of MAE $< s2$ described in Figure \ref{fig:intro} b2).
\subsection{Educated guess of training set size ratios}
While the Bayesian optimization presented in this work reliably identifies the optimal weights between the different levels, it can be computationally expensive, even using a representation of constant size.
A good initial guess for the weights can be obtained from learning curves\cite{Vapnik1998,vapnik1994learningcurves,StatError_Muller1996}.
It has been observed that direct learning curves exhibit a slope close to -1/2 for various representations and different tasks\cite{felix_google, mbdf}.
%TODO Konstantin 03.01.2023 - given the two papers you cite adding "different tasks related to machine learning quantities in chemical space" may be preferrable. Otherwise, now that I think again, it should be good enough.
% TODO Konstantin: (1) The KRR slopes presented in the paper you refer to never reach -1/2, though this might be because they didn't cut off the larger MAEs while calculating the slopes. So this should either be phrased more carefully, or -1/2 should just be stated as the observed result. Does that have a theoretical backing BTW? If you don't know one I can do a quick search for it; I vaguely remember hearing the slope is between -1/4 and -1/2 if your representation is unique. (2) FIG 3 demonstrates that the linear approximation for learning curves breaks down in the beginning; you discuss this in Results & Discussion, but perhaps note here already that the slope is asymptotic?
%TODO Konstantin: If I understand this paper correctly -1/2 is the upper bound of the slope for large enough sample sizes: 10.1287/moor.1080.0317. I'm not sure how transferable the result is to chemical space in general and our datasets in particular, but it might be good enough of a reference for where the -1/2 comes from XD
Therefore, the MAE of direct learning approaches can be estimated from the null model, which represents the MAE when only one data point is given as shown in Figure \ref{fig:intro} b3.
The learning curve of multi-level learning can then be approximated by propagating the uncertainties of these individual models.
In this case, direct learning of $y_\mathrm{HF}$ is combined with learning the differences between HF and MP2, as well as the differences between MP2 and CCSD. The total cost $C$ in this model depends on the number of training points $N$ for different methods of average cost $c$:

\begin{equation}
C = N_\text{HF} \cdot c_\text{HF} + N_\text{MP2} \cdot c_\text{MP2} + N_\text{CCSD} \cdot c_\text{CCSD}
\end{equation}

Meanwhile, the MAE $M$ in the multi-level approach is estimated to emerge from the null model accuracies $m_l$, where $l$ stands for HF, MP2, and CCSD(T), respectively and can be derived as follows using the $Ansatz$ in Figure \ref{fig:intro} b3:
\begin{align}
\mathrm{log}(\mathrm{MAE}) & = -\frac{1}{2} N + \mathrm{log}(m)-\frac{1}{2} \\
 & = -\frac{1}{2}(N-1) + \mathrm{log}(m) \\
 \mathrm{MAE} & = \mathrm{exp}(-\frac{1}{2} (N-1) + \mathrm{log}(m)) \\
 & = m \cdot \mathrm{exp}(-\frac{1}{2}(N-1))
\end{align}
using equation 12 and error propagation for all three levels $l$ leads to the final equation:
\begin{equation}
\mathrm{MAE} = \sqrt{\sum_{l\in \text{HF, MP2, CCSD}}\left(\exp(-0.5 \cdot (\log(N_l) - 1)) \cdot m_l\right)^2}
\end{equation}
Following short code example was used to get the optimal ratio using an average COST estimate $C$ and the null model error $m_l$:
\begin{lstlisting}[language=Python]
import numpy as np
import itertools as it

c  = {"HF": 0.016, "MP": 0.039, "CC": 0.235}
ml = {"HF": 35.7, "MP": 6.16, "CC": 2.76}

def cost_and_error(N_HF, N_MP, N_CC):
    c  = N_HF*c["HF"]+N_MP2*c["MP"]+N_CC*c["CC"]
    eA = ml["HF"]*np.exp(-0.5*(np.log(N_HF)-1))
    eB = ml["MP"]*np.exp(-0.5*(np.log(N_MP)-1))
    eC = ml["CC"]*np.exp(-0.5*(np.log(N_CC)-1))
    return cost, np.sqrt(eA**2+eB**2+eC**2)

def optimal_for_error(thres):
    lvl = [int(_) for _ in 2**np.linspace(0,15)]

    mcost = 1e100
    opt = None
    for NHF,NMP,NCC in it.product(lvl,lvl,lvl):
        cost,error = cost_and_error(NHF,NMP,NCC)
        if cost < mcost and error < thresh:
            mcost = cost
            opt = (NHF, NMP, NCC)
    return opt
\end{lstlisting}
\section{Results and Discussion}

\subsection{Minimizing Training Acquisition Cost within M3L}

\begin{table}[!ht]
\centering
\caption{Summary of all subsets reporting ratios of CCSD(T) equal to 1, MP2, HF, total number of highest level training data ($N^\mathrm{CCSD(T)}$) where each model reaches chemical accuracy (1kcal/mol) and the total CPU time in h needed  (CPU$^\mathrm{tot}$) to train those models. Green cells report best models (lowest training acquisition time). Yellow cells report models where the learning curve could not be extrapolated correctly due to optimizer reach too small training data of the highest levels of theory.}
\label{tab:results}
\small
\begin{tabular}{ll|rr|r|r}
\hline
\multicolumn{2}{c}{Subsets} & MP2 & HF & $N^\mathrm{CCSD(T)}$ & CPU$^\mathrm{tot}$ [h] \\
\hline
\hline
\multirow{6}{*}{QM7b} & Guess & 7 & 68 & 101 & 150.2 \\
& M2L & 4 & 16 & 232 & 151.6 \\
& $\beta$=0 (constr.) & 17 & 67 & 74 & 150.7 \\
& $\beta$=0.25 & 104 & 514 & 107 & \cellcolor{red!25} 1358 \\
& $\beta$=0.50 & 235 & 917 & 47 & \cellcolor{red!25} 1159 \\
& $\beta$=0.75 & 3 & 17 & 243 & \cellcolor{green!25}149 \\
\hline
\multirow{7}{*}{QM9$^\mathrm{LCCSD(T)}$} & Guess & 3 & 35 & 962 & 6398 \\
& M2L & 4 & 16 & 1333 & 6941 \\
& $\beta$=0 (constr.) & 10 & 39 & 813 & 7127 \\
& $\beta$=0.25 & 20 & 100 & 290 & 7690 \\
& $\beta$=0.50 & 5 & 36 & 643 & \cellcolor{green!25} 6346 \\
& $\beta$=0.75 & 10 & 23 & 1018 & 8436 \\
\hline
\multirow{7}{*}{QM9$^\mathrm{CCSD(T)}_\mathrm{AE}$} & Guess & 6 & 162 & 594 & 1337 \\
& M2L & 4 & 16 & 2508 & 3227 \\
& $\beta$=0 (constr.) & 47 & 187 & 413 & 2061 \\
& $\beta$=0.25 & 30 & 411 & 77 & \cellcolor{yellow!25}400 \\
& $\beta$=0.50 & 13 & 160 & 418 &  1130 \\
& $\beta$=0.75 & 10 & 221 & 302 & \cellcolor{green!25} 853 \\
\hline
\multirow{7}{*}{QM9$^\mathrm{CCSD(T)}_\mathrm{EA}$} & Guess & 2 & 23 & 18763 & 12871 \\
& M2L & 4 & 16 & 75163 & 128077 \\
& $\beta$=0 (constr.) & 10 & 35 & 1921 & \cellcolor{green!25}4951 \\
& $\beta$=0.25 & 8 & 252 & 1866 & 9387 \\
& $\beta$=0.50 & 58 & 290 & 684 & 7550 \\
& $\beta$=0.75 & 16 & 49 & 1987 & 7128 \\
\hline
\multirow{7}{*}{EGP} & Guess & 20 & 150 & 370 & 1951 \\
& M2L & 4 & 16 & 6787 & 10382 \\
& $\beta$=0 (constr.) & 6 & 20 & 3529 & 7601 \\
& $\beta$=0.25 & 8 & 80 & 455 & 1535 \\
& $\beta$=0.50 & 5 & 36 & 917 & 1921 \\
& $\beta$=0.75 & 13 & 65 & 268 & \cellcolor{green!25}751 \\
\hline
\hline
\end{tabular}
\end{table}

\begin{figure*}[ht!]
\includegraphics[width=\textwidth]{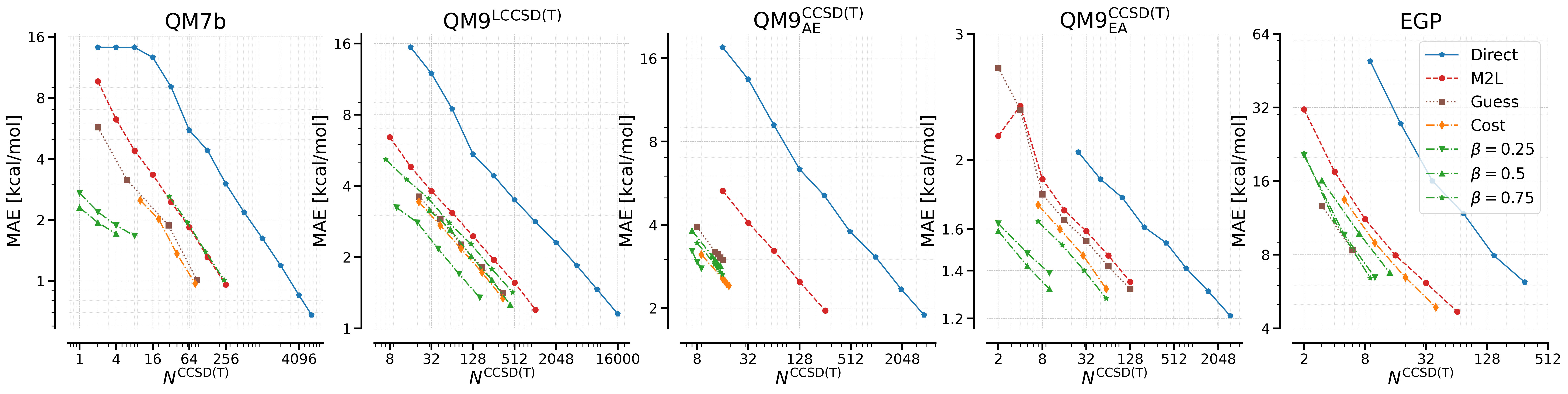}
\caption{Learning curves showing MAE vs. $N^\mathrm{CCSD(T)}$ (highest level of theory) for each subset for direct learning, original work (M2L), educated guess (Guess), Bayesian optimizations (Cost ($\beta=0$) and different $\beta$'s), respectively. }
\label{fig:lc}
\end{figure*}

We optimized the training set ratios at different levels of theory (CCSD(T):MP2:HF) using Bayesian optimization with N$^\mathrm{CCSD(T)}$, N$^\mathrm{MP2}$, and N$^\mathrm{HF}$ as parameters.
We utilized the s2 ratio as initial points for both loss functions $L(\beta)$ (equation \ref{eq:cost}) for $\beta =$ [0.25, 0.5, 0.75], and we also employed the constraint loss function $L(\beta=0) = \mathrm{COST}$ (Figure \ref{fig:intro} b2).
Additionally, we obtained ratios using the "fancy guess" approach leveraging learning theory and the  optimal behaviour of learning curves with a slope of -1/2.

To evaluate the models, we generated learning curves showing the mean absolute error (MAE) vs. the training set size of the highest level of theory ($N^\mathrm{CCSD(T)}$).
Figure \ref{fig:lc} presents learning curves for all subsets. Notably, all models showed improved performance with increasing training set sizes, with CCSD(T) being the highest level, MP2 the intermediate level, and HF the lowest level of theory.
Direct learning exhibited the worst performance, while the heuristic model $s2$ performed slightly better but was outperformed by the optimized models.

Table \ref{tab:results} summarizes the results for all subsets, reporting the ratios MP2 and HF w.r.t CCSD(T) equal to 1, as well as the number of training data for CCSD(T) ($N^\mathrm{CCSD(T)}$) and the total CPU time required for training data acquisition time (CPU$^\mathrm{tot}$).
The models with the lowest training acquisition time are highlighted in green, while models with suboptimal learning curves due to limited training data for the highest level are marked in red.
Those being for QM7b the models $\beta$ = 0.25 and 0.5 with high ratios of MP2 and HF with 1:104:514 for $\beta$ = 0.25 and 1:235:917 for $\beta$ = 0.5, respectively.
In the small training data range of 1 to 8 training points (Figure \ref{fig:lc}) smaller slopes were observed, limiting the slope of the learning curves.
Adding more training data will lead to a similar slopes compared to the other learning curves.
For the subset QM9$^\mathrm{CCSD(T)}_\mathrm{AE}$ the learning curve for $\beta$ = 0.25 has to be analyzed with caution.
Similarly, to QM7b learning curves discussed above, the low data range may not represent the real slope of the learning curve, which again, will converge to the slope of the other learning curves.
Excluding those learning curves, the best models for all subsets are $\beta$ = 0.75, $\beta$ = 0.5, $\beta$ = 0.75, $\beta=0$ (constr.), and $\beta$ = 0.75 for QM7b, QM9$^\mathrm{LCCSD(T)}$, QM9$^\mathrm{CCSD(T)}_\mathrm{AE}$, QM9$^\mathrm{LCCSD(T)}$, QM9$^\mathrm{CCSD(T)}_\mathrm{EA}$, and EGP, respectively.
It is evident that the Bayesian optimization models consistently outperformed other approaches using high $\beta$ values, achieving a reduction in training set size of the highest level and significantly reducing training acquisition time compared to the original work $s2$.
The "fancy guess" approach yielded promising results in a fraction of the time needed by the Bayesian optimization, giving a good first and cheap estimate of optimal ratios.

\begin{figure}[ht!]
\includegraphics[width=0.49\textwidth]{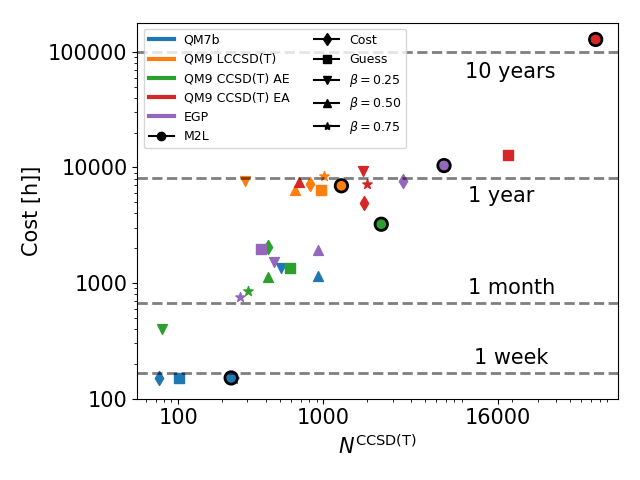}
\caption{Critical points of chemical accuracy on the Pareto front of total training acquisition time vs. training data of the highest level of theory ($N^\mathrm{CCSD(T)}$) for all datasets (QM7b, QM9 LCCSD(T), QM9 CCSD(T) atomization energies and electron affinities, and EGP) where each model reaches chemical accuracy (1 kcal/mol) extrapolated from corresponding learning curves. Original work (M2L) encircled in black.}
\label{fig:criticalpoints}
\end{figure}

To further evaluate the models, critical points for each dataset were analyzed, displaying the training set size of the highest level of theory vs. the total cost (CPU hours) at 1kcal/mol (extrapolated through a linear fit using the slope of the learning curves).
The original work $s2$ was used as a reference point (black circles around points in Figure \ref{fig:criticalpoints}).
Models that show fewer training data needed for smaller datasets (QM7b and QM9$^\mathrm{LCCSD(T)}$) and faster/less accurate calculations are considered more efficient, even though no significant gain in cost was achieved.
On the other hand, for larger datasets like EGP and the electron affinities for QM9, a reduction in the need for high-level training data and significant time savings were observed.
The models marked in yellow in Table \ref{tab:results} with limited training data for the highest level are expected to perform worse due to the inaccurate extrapolation of learning curves. However, with more data, the models' performance is likely to improve, and the total training acquisition time would be further reduced.

\subsection{Density Functional Approximations within M2L}
Density Functional Theory has emerged as a powerful computational tool for studying a wide range of chemical systems due to its advantageous trade-off between efficiency and accuracy.
Key to the success of DFT are the various functionals, each designed with distinct focuses and approximations.
Among these, Generalized Gradient Approximations (GGAs) stand out, leveraging the reduced density gradient to achieve semi-local behavior.
However, GGAs fall short in capturing critical physical aspects, leading to the development of meta-GGAs that aim to include additional quantities like $\tau$ and $\nabla^2\rho$.
Although several meta-GGAs exist, many rely heavily on empirical approaches, resulting in marginal improvements over GGA functionals.

To attain superior accuracy over GGAs, hybrid functionals have proven to be highly effective.
The global hybrid functional takes the form of $E_{X_\sigma}^\mathrm{hybrid} = aE_{X_\sigma}^\mathrm{HF} + (1-a)E_{X_\sigma}^\mathrm{GGA}$, where the precise value of 'a' typically falls in the range of 20-50\%, depending on the functional.
In the realm of Chemistry, hybrid functionals outperform both pure HF and GGA functionals, providing valuable insights into various chemical problems.

However, the use of hybrid functionals necessitates parametrization.
The widely-used B3LYP functional, for instance, contains three parameters represented as $E_{\mathrm{XC}} = aE_{X_\sigma}^\mathrm{HF} + (1-a)E_{X_\sigma}^\mathrm{LSDA} + b\delta E_{X_\sigma}^\mathrm{GGA} + E_{C}^\mathrm{LSDA} + c\delta E_{C}^\mathrm{GGA}$.
While parametrization can enhance performance to some extent, it comes with potential drawbacks, including deleterious effects on numerical sensitivity and oscillatory behavior in the enhancement factor containing the reduced density gradient.
Over-fitting issues, as observed in the Minnesota functionals, also arise in the parametrization process.

Another critical aspect of density functionals is the absence of dispersion characterization in base functionals, impacting the modeling of intermolecular interactions.
To address this, post-SCF corrections are commonly employed, such as D3 corrections \cite{Goerigk2017D3} being a widely used and accurate approach.
\begin{figure*}[ht!]
\includegraphics[width=\textwidth]{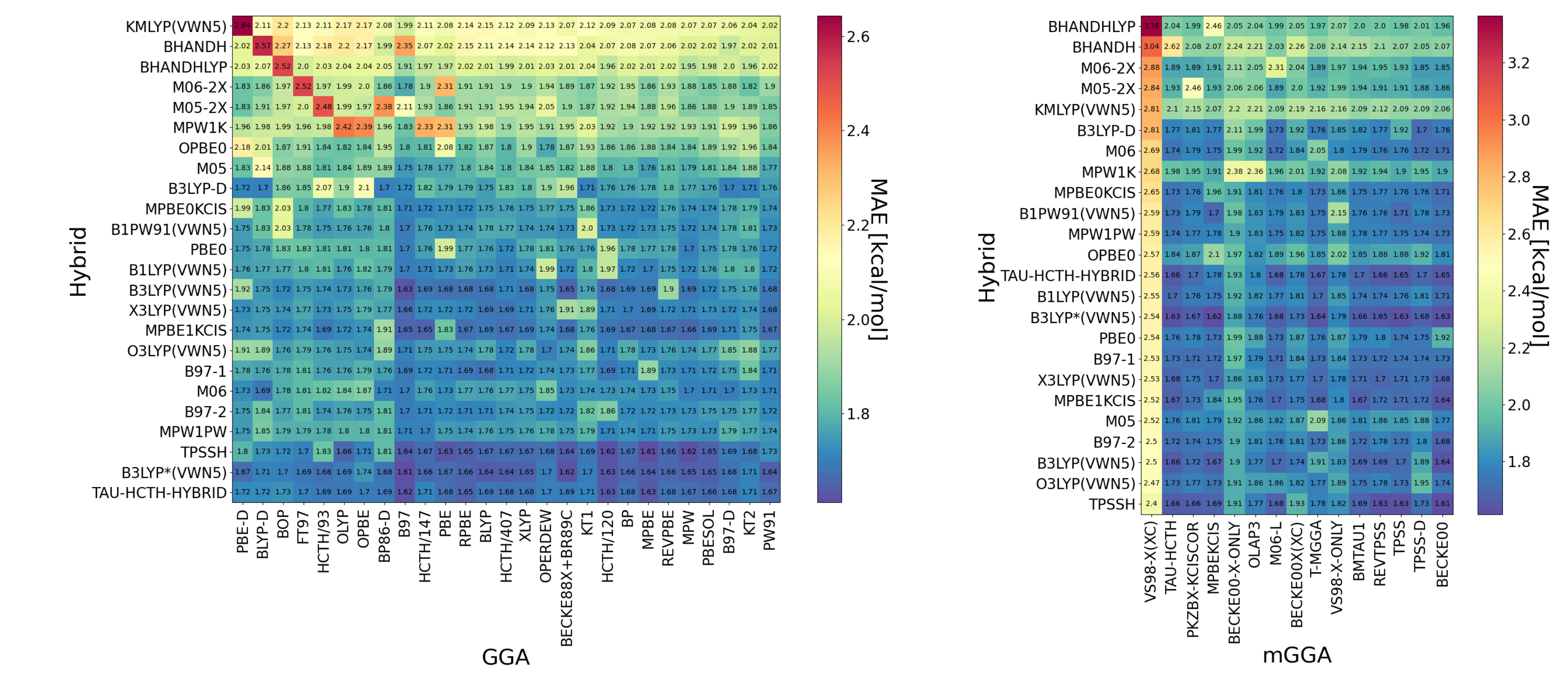}
\caption{MAE for all combinations of LDA$\rightarrow$GGA$\rightarrow$Hybrid (left) and LDA$\rightarrow$mGGA$\rightarrow$Hybrid (right) for QM9 dataset. Rows and columns are sorted according to the mean errors of each column and row, showing no advantage of mGGAs over GGAs in multi-level learning.}
\label{fig:dft}
\end{figure*}

Since there are so many density functionals, with different flavours and parametrizations\cite{Goerigk2017},
we  investigated the hierarchy of Jacob's ladder for the QM9$^\mathrm{DFT}$ dataset containing single-point calculations for 76 functionals per molecule, and using the original M2L ratio $s2$ to train several machine learning models.
Figure \ref{fig:dft} presents the MAE for all combinations of LDA$\rightarrow$GGA$\rightarrow$Hybrid (left) and LDA$\rightarrow$mGGA$\rightarrow$Hybrid (right) for the atomization energies.
The rows and columns in the in Figure \ref{fig:dft} are sorted according to the mean errors of each model, providing insights into the performance of different functionals at each level of theory.

Despite the diversity of density functionals, the results showed that there was no significant gain in using mGGA over GGA, considering the higher calculation timings without any clear benefit.
Notably, some functionals, such as PBE-D and VS98-X(XC), performed poorly overall in both GGA and mGGA cases.
This could be attributed to the missing dispersion correction in most hybrid functionals for PBE-D.
On the other hand, B97-D, KT2, and PW91 emerged as the best performing functionals for the GGA case, while BECKE00, TPSS-D, and TPSS were the top performers for the mGGA case.

Interestingly, B3LYP flavors consistently reached the top three positions among the hybrid functionals for both GGA and mGGA cases.
In contrast, KMLYP and BHANDH flavors underperformed in both cases, rendering them unsuitable for multi-level learning approaches.
Furthermore, combining B97 with the best-performing hybrid functionals yielded the lowest errors in the GGA case.

In conclusion, density functionals exhibit varying performance levels, and the combination of different datasets to lift a target dataset onto a higher level of theory could be a viable approach in real-world applications.
Although mGGA functionals did not provide significant advantages over GGA functionals, multi-level learning with different steps on Jacob's ladder remains feasible, especially considering the widespread use of density functional theory in machine learning applications.

Overall, our findings demonstrate the potential for significant reductions in training acquisition cost using Bayesian optimization and the efficacy of multi-level learning with density functional theory in quantum chemistry applications.

\section{Conclusion}
%
% add most important results
In this study, we introduced M3L as a means to minimize training acquisition cost for machine learning models in quantum chemistry applications.
Through Bayesian optimization, we optimized the ratios of quantum chemistry levels (CCSD(T):MP2:HF) to achieve the most efficient learning curves.
Our M3L results demonstrate the effectiveness of this approach in reducing the training set size of the highest level of theory and significantly reduce training acquisition costs by up to 18 times compared to the original heuristic M2L approach ($s2$).

The Bayesian optimization models consistently outperformed other approaches, providing better ratios and resulting in more efficient learning curves.
Moreover, the Educated Guess approach showed promise, achieving similar performance to the Bayesian optimization models but with a fraction of the cost.
However, further exploration and comparison on larger datasets are warranted to fully evaluate its potential.

Additionally, we investigated the hierarchy of Jacob's ladder for density functional theory (DFT) using the QM9$^\mathrm{DFT}$ dataset.
Surprisingly, we found no significant advantage in using meta-GGA (mGGA) functionals over generalized gradient approximation (GGA) functionals, despite the higher complexity (including more physics) for mGGA.
This suggests that for multi-level learning with DFT, GGA functionals might be more efficient without compromising accuracy.
Furthermore, we identified the best-performing functionals for both GGA and mGGA cases, highlighting the importance of choosing suitable functionals for specific applications.

\section{Supplementary Information}
All data, scripts, and results will be comprehensively documented within the final publication. Should there be a need, interested parties are welcome to request these materials directly from the authors.
%All data generated or used in this study is available at \cite{SI}.
%This file contains several directories: \\
%(i) \textit{geometries}: The geometries used in this study for all five subsets (QM7b, QM9$^\mathrm{LCCSD(T)}$, QM9$^\mathrm{CCSD(T)}_\mathrm{AE/EA}$, and EGP) \\
%(ii) \textit{scripts}: containing example scripts for KRR, Bayesian optimization, educated guess, and plotting/fitting scripts\\
%(iii) \textit{CSVs}: containing CSV files with properties and timings of the five subsets mentioned in (i), as well as the results from the KRR predictions from learning curves and the grid scan for the DFT data.\\
%Following scheme shows the directory structure:
%
%\begin{forest}
%  for tree={
%    font=\ttfamily,
%    grow'=0,
%    child anchor=west,
%    parent anchor=south,
%    anchor=west,
%    calign=first,
%    inner xsep=7pt,
%    edge path={
%      \noexpand\path [draw, \forestoption{edge}]
%      (!u.south west) +(7.5pt,0) |- (.child anchor) pic {folder} \forestoption{edge label};
%    },
%    before typesetting nodes={
%      if n=1
%        {insert before={[,phantom]}}
%        {}
%    },
%    fit=band,
%    before computing xy={l=15pt},
%  }  
%[SI
%  [geometries
%   [xyz\_qm7b
%   ]
%   [xyz\_qm9\_lccsd
%   ]
%   [xyz\_qm9\_ccsd\_ae\_ea
%   ]
%   [xyz\_egp
%   ]
%  ]
%  [scripts
%  ]
%  [CSVs
%   [DFT
%   ]
%   [learning\_curves
%   ]
%   [properties
%   ]
%  ]
%]
%\end{forest}

\section*{Acknowledgement}
The author of this manuscript utilized Chat-GPT3.5. They assume complete responsibility for every sentence presented in the document.
O.A.v.L. has received funding from the European Research Council (ERC) under the European Union’s Horizon 2020 research and innovation programme (grant agreement No.~772834).
This project has received funding from the European Union’s Horizon 2020 research and innovation programme under grant agreement No.~957189.
This research was undertaken thanks in part to funding provided to the University of Toronto's Acceleration Consortium from the Canada First Research Excellence Fund (CFREF).
O.A.v.L. has received support as the Ed Clark Chair of Advanced Materials and as a Canada CIFAR AI Chair.  
This result only reflects the
author's view and the EU is not responsible for any use that may be made of the
information it contains.
The project has been supported by the Swedish Research Council (Vetenskapsrådet), and the project the prothe Swedish National Strategic e-Science program eSSENCE as well as by computing resources from the Swedish National Infrastructure for Computing (SNIC/NAISS).
\bibliography{literature}
\end{document}